\documentstyle[aps,prl,preprint,floats,epsfig]{revtex}
\begin{document}
\draft

\preprint{\tighten\vbox{\hbox{\hfil CLNS 97/1516}
                        \hbox{\hfil CLEO 97-23}
}}

\title{Flavor-Specific Inclusive $B$ Decays to Charm}

\author{CLEO Collaboration }
\date{October 23, 1997}
 
\maketitle
\tighten

\begin{abstract}

We have measured the branching fractions for
$B \rightarrow \bar D X$, $B \rightarrow D X$,
and $B \rightarrow \bar D X \ell^+ \nu$, where ``$B$'' is an average over $B^0$
and $B^+$, ``$D$'' is a sum over $D^0$ and $D^+$, and
``$\bar D$'' is a sum over $\bar D^0$ and $D^-$.
From these results and some previously measured branching fractions,
we obtain ${\cal B}(b \rightarrow c \bar c s)$ = (21.9 $\pm$ 3.7)\%,
${\cal B}(b \rightarrow s g) <$  6.8\% @ 90\% c.l, and 
${\cal B}(D^0 \rightarrow K^- \pi^+)$ = (3.69 $\pm$ 0.20)\%.
Implications for the ``$B$ semileptonic decay problem'' (measured branching
fraction being below theoretical expectations) are discussed.  The increase in
the value of ${\cal B}(b \rightarrow c \bar c s)$ due to $B \rightarrow D X$
eliminates 40\% of the discrepancy.

 \end{abstract}


\newpage

{
{\renewcommand{\thefootnote}{\fnsymbol{footnote}}

\begin{center}
T.~E.~Coan,$^{1}$ V.~Fadeyev,$^{1}$ I.~Korolkov,$^{1}$
Y.~Maravin,$^{1}$ I.~Narsky,$^{1}$ V.~Shelkov,$^{1}$
J.~Staeck,$^{1}$ R.~Stroynowski,$^{1}$ I.~Volobouev,$^{1}$
J.~Ye,$^{1}$
% ...
\end{center}
 
\small
\begin{center}
$^{1}${Southern Methodist University, Dallas, Texas 75275}\\
% ...
$^{13}${University of Hawaii at Manoa, Honolulu, Hawaii 96822}\\
$^{14}${University of Illinois, Urbana-Champaign, Illinois 61801}\\
% ...
$^{25}${Stanford Linear Accelerator Center, Stanford University, Stanford,
California 94309}
\end{center}

\setcounter{footnote}{0}
}
\newpage


\section{Introduction}

There has been a longstanding problem in heavy flavor physics of the measured
$B$ semileptonic decay branching fraction\cite{CLEO_SL} being smaller than
theoretical expectations\cite{Bigi,Hitoshi-etal}.  One possible
explanation\cite{Bigi} is
a larger-than-expected flavor-changing neutral current (FCNC) contribution, due
to new physics. Another\cite{Hitoshi-etal} is an enhanced rate for
$b \rightarrow c \bar c s^\prime$ ($s^\prime$ denotes the weak isospin
partner of $c$). An argument against an enhanced
$b \rightarrow c \bar c s^\prime$ rate is that it would conflict with the
measured branching fraction for $B \rightarrow \bar D X$ plus
$B \rightarrow D X$.  That measurement relies on a knowledge of
${\cal B}(D^0 \rightarrow K^- \pi^+)$, however, and if
that is in error, the measurement of the branching fraction of $B$ to charm or
anticharm will be also.  We address all three issues by measuring the yields
of the flavor-specific inclusive $B$ decay processes
$B \rightarrow D X$, $B \rightarrow \bar D X$, and
$B \rightarrow \bar D X \ell^+ \nu$ in a sample of $B \bar B$ events in which
at least one
$B$ decays semileptonically.  (Herein, ``$B$'' represents an average over
$B^0$ and $B^+$, ``$D$'' a sum over $D^0$ and $D^+$, and ``$\bar D$'' a sum over
$\bar D^0$ and $D^-$\cite{conjugate}.  We use the term ``upper vertex $D$'' for
a $\bar D$ produced from the charm quark from $W \rightarrow \bar c s$, and
``lower vertex $D$'' for a $D$ produced from the charm quark from $b \rightarrow
c$.)

These yields, and ratios among them, provide information on the above-mentioned
issues as follows:

(i) The fraction of semileptonic $B$ decays that proceed through
$B \rightarrow \bar D X \ell^+ \nu$, $f_{SL}$, differs from 100\% only because
of small contributions from
$b \rightarrow u \ell \nu$ and $B \rightarrow D_s^- K X \ell^+ \nu$
(``lower vertex $D_s$'').  The measured fraction is inversely proportional to
the assumed $D$ absolute branching fraction (in our case
${\cal B}(D^0 \rightarrow K^- \pi^+)$)
and scaling the yield to agree with expectations gives a new method to measure
that branching fraction.

(ii) The fraction of all $B$ decays that proceed through
$B \rightarrow \bar D X$, $f_{all}$,  differs from 100\% because of
$b \rightarrow u$ decays, lower vertex $D_s$, formation of $c \bar c$ bound
states, formation  of charmed baryons, and
FCNC processes such as $b \rightarrow s g$,
$b \rightarrow d g$, $b \rightarrow s q \bar q$, $b \rightarrow d q \bar q$
(which we will refer to collectively as ``$b \rightarrow s g$'').  As all
processes except $b \rightarrow s g$ have been measured, the ratio
$f_{all}/f_{SL}$ provides a measurement of the branching fraction for 
$b \rightarrow s g$.  By taking
the ratio of $f_{all}$ to $f_{SL}$, rather than just using $f_{all}$, we
eliminate dependence on the $D^0 \rightarrow K^- \pi^+$ branching ratio, and
reduce dependence on $D$ detection efficiency.

(iii) The process $B \rightarrow D X$ proceeds via the the quark-level process
$b \rightarrow c \bar c s^\prime$, and thus the ratio of the yields for
$B \rightarrow D X$ and $B \rightarrow \bar D X$, i.e., ratio of upper to lower
vertex charm, provides information on the
rate of that process relative to $b \rightarrow c \bar u d^\prime$.

The typical inclusive $B$ decay branching fraction measurement 
averages over $B$ and $\bar B$ initial states for a given final
state, and consequently averages over particle and antiparticle final states
for a given initial state ($B$ or $\bar B$), losing the flavor-specific
information sought here. In 1987 CLEO
developed a technique for measuring inclusive $B$ decay branching fractions
separately to particle and antiparticle final states, and applied it to
inclusive kaon decays\cite{Tipton-thesis,kaon-lepton-PRL}. Here we apply
similar techniques to inclusive charm decays.

The principle underlying the 1987 technique is that if one $B$ from a
$B \bar B$ pair from the $\Upsilon$(4S) decays semileptonically,
with a high momentum lepton, then the other decay products from that $B$
will have substantial angular correlations with the lepton, tending to come
off back-to-back to it, while the decay products from the other $B$ have
negligible angular correlations with the lepton. The lepton tags the flavor
of its parent $B$, and thus also the other $B$ (with a correction
needed for mixing). By plotting the distribution in the angle between
$D \ell^+$ (and $\bar D \ell^-$) pairs, and separately the distribution in
the angle between $D \ell^-$ (and $\bar D \ell^+$) pairs, and extracting an
isotropic component and a peaking component from each, yields are obtained
for four processes: $B \rightarrow \bar D X \ell^+ \nu$,
$B \rightarrow  D X \ell^+ \nu$, $B \rightarrow \bar D X$, and
$B \rightarrow D X$.  Of these, $B \rightarrow  D X \ell^+ \nu$ should be zero.

For low $D$ momenta, the technique just described loses statistical power and
becomes sensitive to the shape assumed in fitting for the peaking component.
(In the limit
that the $D$ momentum vanishes, the $D$--lepton angular correlation clearly
contains no information.)  Consequently, we have developed a second technique,
based on charge correlations alone.  We measure three yields:
the number of $D\ell^-$ (and $\bar D\ell^+$) pairs,  equal to the sum of
$B \rightarrow \bar D X \ell^+ \nu$ and $B \rightarrow D X$ yields in the
lepton-tagged data sample; the number of $D\ell^+$ (and
$\bar D\ell^-$) pairs,  equal to the sum of $B \rightarrow D X \ell^+ \nu$ and
$B \rightarrow \bar D X$ yields in the lepton-tagged sample;
and the number of $D$ (and $\bar D$) mesons in an untagged sample, equal to the
sum of $B \rightarrow \bar D X$ and $B \rightarrow D X$ yields in the untagged
sample.  Using the fact that the rate for $B \rightarrow  D X \ell^+ \nu$
vanishes, and scaling the last-mentioned yield by the ratio of the sizes of the
tagged and
untagged data samples, these yields give the yields for the
other three processes: $B \rightarrow \bar D X \ell^+ \nu$,
$B \rightarrow \bar D X$, and $B \rightarrow D X$.
Using a combination of the angular correlation and charge correlation
techniques, we have obtained those three yields for the sum of
$D^0$ and $D^+$ mesons.

\section{Procedures}

The data were taken with the CLEO detector\cite{Kubota} at the Cornell Electron
Storage Ring (CESR), and consist of 3.2 fb$^{-1}$ on the $\Upsilon$(4S)
resonance and 1.6 fb$^{-1}$ at a center-of-mass energy 60 MeV below the
resonance. The on-resonance sample contains 3.3 million $B \bar B$ events and
10 million continuum events. The CLEO detector measures charged particles
over 95\% of $4 \pi$ steradians with a system of cylindrical drift chambers.
Its barrel and endcap CsI electromagnetic calorimeters cover 98\% of $4 \pi$.
Hadron identification is provided by specific ionization ($dE/dx$) measurements
in the outermost drift chamber and by time-of-flight counters (ToF).  Muons are
identified by their ability to penetrate iron; electrons by $dE/dx$,
comparison of track momentum with calorimeter cluster energy, and
track/cluster position matching.

We select hadronic events containing at least 4 charged tracks.  We require a
value of the ratio  of Fox-Wolfram parameters\cite{Fox-Wolfram}
$R_2 \equiv H_2/H_0 < 0.5$, to
suppress continuum events.  Events containing at least one lepton with momentum
between 1.5 and 2.8 GeV/c and surviving a $\psi \rightarrow \ell^+ \ell^-$ veto
are scanned for $D^0$, $D^+$, and charge conjugates.  (For the untagged sample,
we drop the lepton requirement.)  We detect
$D^0$ and $D^+$ via the $K^- \pi^+$ and $K^- \pi^+ \pi^+$
decay mode, respectively.  Tracks used
as candidate $D$ decay products must have
$dE/dx$ and/or ToF values within $2\sigma$ of expectations for the particle
assignment made ($K$ or $\pi$).  For $D^0 \rightarrow K^- \pi^+$, particle
identification must rule out the $\bar D^0 \rightarrow \pi^- K^+$ option.

For candidate $D$'s, we histogram the $K \pi$ ($K \pi \pi$) mass
for four intervals in $\cos \theta_{D-\ell}$ and
four intervals in $D$ momentum, separately for the two charge correlations with
the lepton.  These 64 mass distributions are
fit to double-Gaussian signal peaks and polynomial backgrounds, to extract $D$
yields.  These are corrected for detection
efficiency, determined by Monte Carlo simulation augmented by studies of
particle ID efficiency that use data (a sample of $D^{*+} \rightarrow D^0 \pi^+,
\ D^0 \rightarrow K^- \pi^+$ events).  
We perform small subtractions for continuum background (using
below-$\Upsilon$(4S)-resonance data) and for hadrons misidentified as leptons
(using hadrons in place of leptons and weighting by the ``faking probability'').
Small  corrections are made to the $D^0$ yields for the
singly-Cabibbo-suppressed decays $D^0/\bar D^0 \rightarrow K^- K^+$ and
$D^0/\bar D^0 \rightarrow \pi^- \pi^+$ which combine with a single failure of
particle ID to make satellite peaks, for the doubly-Cabibbo-suppressed decay
$D^0 \rightarrow K^+ \pi^-$\cite{CLEO_DCSD}, and for double failures of particle
ID, with $\pi^- K^+ $ treated as $K^- \pi^+$.  A small correction is made to
$D^+$ yields for the decay $D_s^+ \rightarrow K^- K^+ \pi^+$ with the $K^+$
misidentified as a $\pi^+$.  

The $D$ yields for each momentum interval, charge correlation, and $D$ type, are
histogrammed vs. $\cos \theta_{D-\ell}$, 16 distributions in all.  For the high
$D$ momentum intervals 1.3 -- 1.95 and 1.95 --2.6 GeV/c, we fit the $\ell^- D$
angular distributions to an isotropic component and a backward-peaking
component, with fitting functions obtained from Monte Carlo simulation.  We fit
the $\ell^+ D$ angular distributions to an isotropic component alone.  For the
low $D$ momentum intervals 0.0 -- 0.65 and 0.65 -- 1.3 GeV/c, we use the charge
correlation technique, summing over $\cos \theta_{D-\ell}$.  We sum the yields
so obtained over $D$ momentum intervals, and over charged and neutral $D$'s,
correcting for $D^0$ and $D^\pm$ branching fractions, using
${\cal B}(D^0 \rightarrow K^- \pi^+)$ = 3.91\%\cite{CLEO_DKPI}, and
${\cal B}(D^+ \rightarrow K^- \pi^+ \pi^+)/{\cal B}(D^0 \rightarrow K^- \pi^+)$
= 2.35\cite{CLEO_DKPIPI}.  We obtain yields for $D$ and lepton from the same
$B$, and from different $B$'s, as follows.
$N(D \ell^- + \bar D \ell^+, {\rm same} B)
= (3.75 \pm 0.11)\times 10^5$, $N(D \ell^- + \bar D \ell^+,$ different
$B$'s)$
= (6.66 \pm 0.77)\times 10^4$, and $N(D \ell^+ + \bar D \ell^-,$ different
$B$'s)$ = (3.18 \pm 0.08)\times 10^5$, in a sample containing $4.24 \times 10^5$
leptons.  For illustrative purposes, we show
$\cos \theta_{D-\ell}$ distributions summed over momentum intervals and over
$D^0$ and $D^+$, as Fig.~1.  The $\ell^- D + \ell^+ \bar D$ distribution shows
strong back-to-back peaking from $B \rightarrow \bar D X \ell^+ \nu$, while the
$\ell^- \bar D + \ell^+ D$ shows no such peaking, due to the nonexistence of
$B \rightarrow D X \ell^+ \nu$.  One also notes a much larger
isotropic component in $\ell^- \bar D + \ell^+ D$, because of the large rate for
$B \rightarrow \bar D X$ and a small rate for $B \rightarrow D X$ (and a small
rate for mixing $B^0 \rightarrow \bar B^0 \rightarrow D X$).  

If the lepton and $D$ come from the same $B$, then the lepton tags {\it that}
$B$ correctly.  The lepton can't be from decay of $D$, because the $D$ was
{\it detected} via a hadronic decay mode.  It can't be from $\psi$, because the
rate for $B \rightarrow \psi \bar D X$ is negligible.  If there are two $D$'s
from the same $B$, leptons from either will be below our 1.5 GeV/c momentum cut.
If the $B$ has mixed,
nonetheless the lepton correctly tags the $b$ flavor at the instant of decay,
which is what is relevant for understanding the $D$ from the same $B$.  But, if
the lepton and $D$ come from different $B$'s, then the tagging of {\it both}
$B$'s is now imperfect: the ancestor of the lepton because leptons from charm
decay and leptons from $\psi$ now contribute; and the ancestor of the $D$ for
those reasons and in addition because of $B^0 - \bar B^0$ mixing.  Corrections
are thus required when using the yields involving lepton and $D$ from different
$B$'s. These corrections depend on $f_m$ (the probability that a lepton mistags
its ancestor $B$), and $\chi$ (the mixing parameter).

\section{Results and Interpretation}

We extract three distinct pieces of physics from the three yields given above.
For each, we have considered systematic errors due to uncertainties in each of
the previously-mentioned corrections, uncertainties from fitting mass peaks and
$\cos \theta_{D-\ell}$ distributions, and uncertainties in efficiency and $D$
branching fractions.

(i) First consider $\Gamma( B \rightarrow D X)/
\Gamma( B \rightarrow \bar D X)$, the ratio of ``upper vertex'' charm
to ``lower vertex'' charm.  This ratio $U/L$ is obtained from
$x = N(D \ell^- + \bar D \ell^+,$ different $B$'s$)/ N(D \ell^+ + \bar D
\ell^-,$ different $B$'s$)$, by correcting for mixing and mistags.
$U/L = (x - F_m)/(1 - x F_m)$, where
$F_m = (f_m + f^{\prime})/(2 - f_m - f^{\prime})$, and
$f^{\prime} = f_m + \chi - 2 f_m \chi$.  We use $\chi$ = 0.157 as measured by
CLEO with dileptons\cite{Mixing-PRL}, and $f_m = 0.027$ as found there, thereby
achieving cancellation of some systematic errors in $F_m$, giving
$F_m = 0.112 \pm 0.011$. From the yields given above, $x = 0.210 \pm 0.025$,
leading to
\begin{equation}
{\Gamma(B \rightarrow D X)\over
\Gamma(B \rightarrow \bar D X)} = 0.100 \pm 0.026 \pm 0.016,
\end{equation}
where the first error is statistical and the second is systematic, dominated by
the uncertainties in mixing correction($\pm$ 0.012) and the $\cos \theta_{D -
\ell}$ fitting function ($\pm$ 0.008).  This result is surprisingly large, as
conventional wisdom held that $b \rightarrow c \bar c s$ would hadronize
dominantly into $D_s$.  However, Buchalla {\it et al.}\cite{Hitoshi-etal} have
argued that the $D^0$, $D^+$ component should be substantial.

In Fig.~2 we plot the momentum distribution of these ``upper vertex'' $D^0$,
$D^+$, obtained by applying the analysis just described to each of the four $D$
momentum bins.  The spectrum is softer than that for ``lower vertex'' $D$'s,
also shown.  It is well described by 3-body $D^{(*)} D^{(*)} K^{(*)}$
phase space, if one allows one or two of the particles to be the vector states.
CLEO has observed such decay modes\cite{Roy}.

(ii) Next consider the fraction of all $B$ decays to $\bar D$, $f_{all}$,
divided by the fraction  of semileptonic $B$ decays to $\bar D$, $f_{SL}$,
i.e., the double-ratio of widths ${\Gamma (B \rightarrow \bar D X)\over
\Gamma (B \rightarrow all)} /{\Gamma (B \rightarrow \bar D X \ell^+ \nu) \over
\Gamma (B \rightarrow X \ell^+ \nu)}$ .  We obtain this from the ratio of
yields
$N(D \ell^+ + \bar D \ell^-,$ different $B$'s$)/N(D \ell^- + \bar D \ell^+,
{\rm same} B) \equiv z_{raw}$.
Corrections are required to the ``different $B$'s'' yield for mixing and
mistags.
Also, leptons from unvetoed $\psi$ and from secondary decays (3.3 $\pm$ 0.7\%
of all leptons) don't contribute to the peaking yield, and so a correction is
required for that, leading to
$z_{cor} = 0.967 z_{raw}
/ [(1 - 0.5 f_m - 0.5 f^{\prime })(1 + F_mU/L)]$, where $U/L = 0.100$, as
found above.  Applying all corrections, we have
\begin{equation}
{\Gamma (B \rightarrow \bar D X)\over
\Gamma (B \rightarrow all)} /{\Gamma (B \rightarrow \bar D X \ell^+ \nu) \over
\Gamma (B \rightarrow X \ell^+ \nu)} \equiv f_{all}/f_{SL}
 = 0.901 \pm 0.034 \pm 0.015 .
\end{equation}
One expects both $f_{all}$
and $f_{SL}$
to be close to 1.0.  The first ratio will
be less than 1.0 because of $b \rightarrow u$ transitions
($2\vert V_{ub}/V_{cb} \vert^2$, where the 2 is a phase space factor),
lower vertex $D_s$ (2\%), bound $c \bar c$ states
(3.0 $\pm$ 0.5\%\cite{CLEO_charmonium}),
baryons (6.5 $\pm$ 1.5\%\cite{CLEO_baryons}), and
$b \rightarrow s g$ (to be extracted).  The second ratio will be less than 1.0
because of $b \rightarrow u$ transitions
($3\vert V_{ub}/V_{cb} \vert^2$, enhanced by the 1.5 GeV/c lepton momentum
requirement), and lower vertex $D_s$ (1.0 $\pm$ 0.5\%, suppressed by the lepton
momentum requirement).  These lead to


\begin{eqnarray}
f_{all}/ f_{SL} = &  1.0 + \vert V_{ub}/V_{cb} \vert^2
- (0.010 \pm 0.005) - (0.030 \pm 0.005)
 - (0.065 \pm 0.015) - {\cal B}(b \rightarrow s g)
\end{eqnarray}
Here $b \rightarrow s g$ is symbolic for all FCNC processes.  Using
$\vert V_{ub}/V_{cb} \vert^2 = 0.008 \pm 0.003$, we obtain
${\cal B}(b \rightarrow s g) = (0.2 \pm 3.4 \pm 1.5 \pm 1.7)$\%, where the first
error is statistical, the second systematic on $z$, and the third the
uncertainties in  Expression (3).  From this we obtain an
upper limit ${\cal B}(b \rightarrow s g) < 6.8$\%, @ 90\% C.L.
The dominant components of the systematic error on $z$ are from mixing
($\pm$1.2\%) and unvetoed and secondary leptons ($\pm$ 0.6\%).

(iii) Finally consider the fraction of semileptonic $B$ decays to $\bar D^0$ or
$D^-$, i.e., $f_{SL} \equiv \Gamma(B \rightarrow \bar D X \ell^+ \nu)/
\Gamma(B \rightarrow X \ell^+ \nu)$.  We obtain this fraction by dividing the
yield $N(D \ell^- + \bar D \ell^+, {\rm same} B)$ by the number of
leptons from $B$ semileptonic decay, 96.7\% of the total of $4.24 \times 10^5$
leptons in our sample.  We find
$0.914 \pm 0.027 \pm 0.042$.  This
number is inversely proportional to the value used for ${\cal B}(D^0 \rightarrow
K^- \pi^+)$. The expected value of the ratio of widths
is $\Gamma(B \rightarrow \bar D X \ell^+ \nu)/
\Gamma(B \rightarrow X \ell^+ \nu) = 1.0 - 3\vert
V_{ub}/V_{cb}\vert^2 - 0.010 \pm 0.005 {\rm (for} \bar B \rightarrow D_s^+ K X
\ell^- \nu)$.  Taking $3\vert V_{ub}/V_{cb} \vert^2 = 0.023 \pm 0.008$, we find
the expected ratio of widths to be 0.968 $\pm$ 0.010, differing from the
measured value by one standard deviation.  We set measured and expected values
of the ratio equal to each
other, and solve for the $D^0$ branching fraction, finding
${\cal B}(D^0 \rightarrow K^- \pi^+) = (3.69 \pm 0.11 \pm 0.16 \pm 0.04)\%$,
where the first error is statistical, the second systematic in the measured
ratio, and the third systematic in the predicted ratio.  The dominant systematic
errors are from uncertainties in $D$ detection efficiency ($\pm$ 0.10\%), mass
peak fitting ($\pm$ 0.09\%), and the ratio of $D^+$ to $D^0$ branching ratios
($\pm$ 0.08\%).
This value for the
branching fraction, (3.69 $\pm$ 0.20)\%, is to be compared with recent
measurements by CLEO of (3.91 $\pm$ 0.19)\%\cite{CLEO_DKPI}
and (3.81 $\pm$ 0.22)\%\cite{Alexey}, by ALEPH of
(3.90 $\pm$ 0.15)\%\cite{aleph} and the PDG value of
(3.83 $\pm$ 0.12)\%\cite{PDG96}.

\section{The $B$ Semileptonic Decay Branching Fraction Problem}

In Table~I we list all the components of $B$ decay, give their branching
fractions
(based on measurement or theory), and see if they sum to 100\%.  We express some
in terms of $b_{SL}$, the $B$ semileptonic decay branching fraction, for which
we use\cite{CLEO_SL} (10.49 $\pm$ 0.46)\%.  The factor of 0.25 for
$b \rightarrow (c\ {\rm or}\ u) \tau \nu$ is a phase space factor.  The factor
$r_{ud}$ for $b \rightarrow (c\ {\rm or}\ u) u d^\prime$ would be 3 from color
counting, but with QCD
corrections\cite{Bagan-QCD} is 4.0 $\pm$ 0.4.  This analysis has two pieces of
information to add to Table~I.  First, the upper vertex $\bar D^0, D^-$
contribution of (7.9 $\pm$ 2.2)\% is obtained from our measured value of
$\Gamma(B \rightarrow D^0 \,{\rm or} \,D^+ X)/
\Gamma(B \rightarrow \bar D^0 \,{\rm or} \,D^- X)$ combined with the rate for
inclusive $D^0 + D^+$ (63.6\% + 23.5\%)\CITE{Moneti}, and leads to a branching
fraction for $b \rightarrow (c \,{\rm or} \, u) \bar c s^\prime$ of (21.9 $\pm$
3.7)\%.
Second, we have a value
(with large errors) for the FCNC term.  One sees that the upper vertex
$\bar D^0, D^-$ contribution accounts for close to half of the shortfall of the
sum of all modes from unity.  The
remaining shortfall is less than 2 standard deviations.  If we adjust $r_{ud}$
to bring the sum to 100\%, we find $r_{ud}$ = 5.2 $\pm$ 0.6.

\section{Acknowledgments}

We thank Isard Dunietz for many informative conversations.
We gratefully acknowledge the effort of the CESR staff in providing us with
excellent luminosity and running conditions.
This work was supported by 
the National Science Foundation,
the U.S. Department of Energy,
the Heisenberg Foundation,  
the Alexander von Humboldt Stiftung,
Research Corporation,
the Natural Sciences and Engineering Research Council of Canada,
the A.P. Sloan Foundation,
the Swiss National Science Foundation,
and the Yonsei University faculty research fund.

\begin{figure}
\centering \leavevmode
\epsfxsize=6in \epsfbox{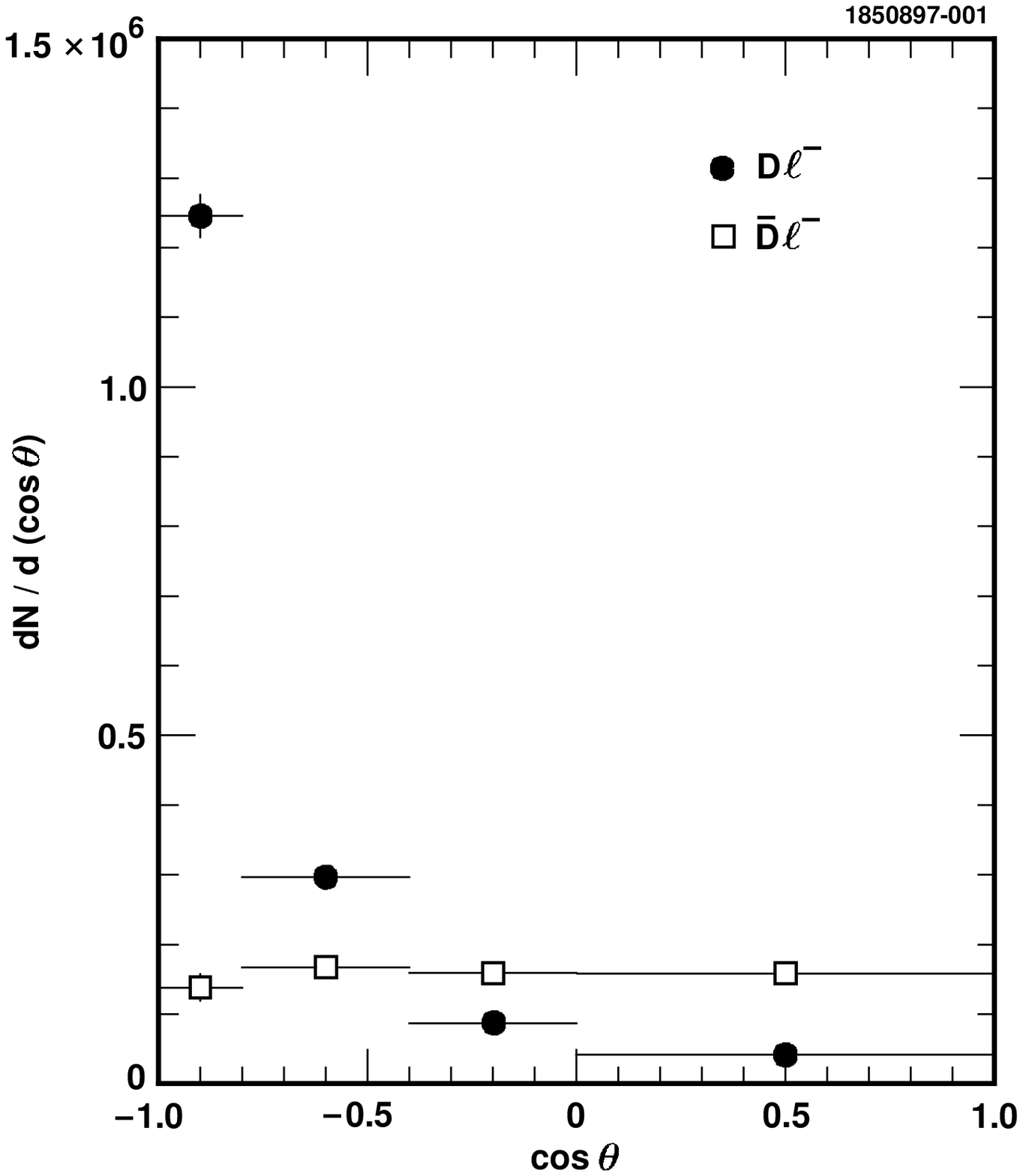} \hfill
\caption{Yield of $D \ell$ events vs $\cos \theta_{D-\ell}$. 
$ D^0 \ell^-\  +\ D^+ \ell^-\ $ plus charge conjugate, summed
over $D$ momentum are shown as solid circles, while
$ \bar D^0 \ell^-\ +\ D^- \ell^-$\ plus charge conjugate,
summed over $D$ momentum are shown as open squares.}
\hfill
\end{figure}

\begin{figure}
\centering \leavevmode
\epsfxsize=6in \epsfbox{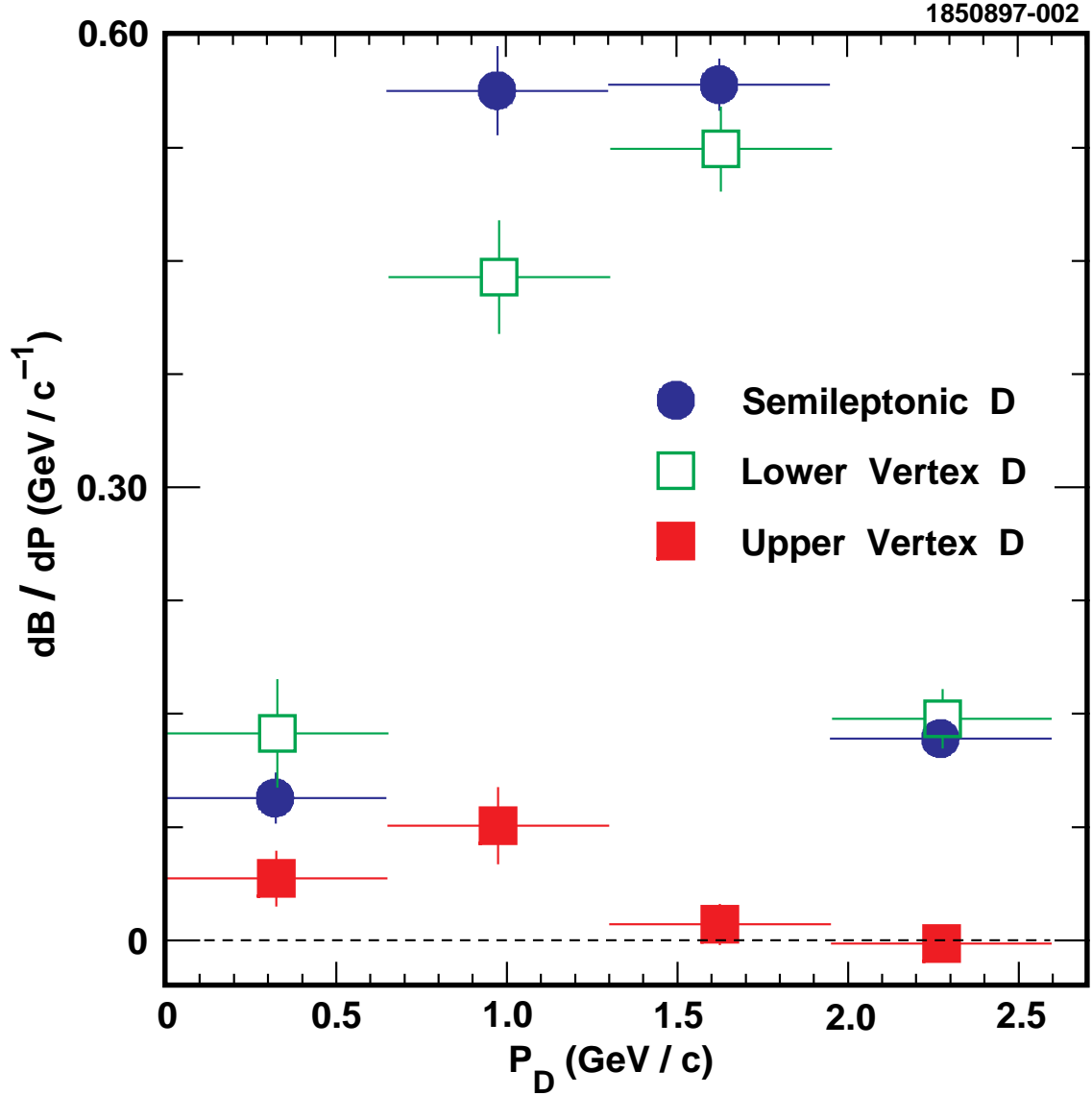} \hfill
\caption{$D$ momentum distributions.
``Upper vertex'' $D^0 + D^+$, i.e., from $B \rightarrow D X$ are shown as
solid squares, while 
``lower vertex'' $D^0 + D^+$, from $\bar B \rightarrow D X$ are shown as
open squares and
``lower vertex'' $D^0 + D^+$, from 
$\bar B \rightarrow D X \ell \nu$ are shown as solid circles.  Vertical scale
gives branching fraction per unit momentum, for upper and lower vertex $D$'s,
and same divided by total semileptonic decay branching fraction for semileptonic
$D$'s.}
\hfill
\end{figure}

\pagebreak

\tighten

\begin{table}[h]
\begin{center}
\caption{All components of $B$ decay, with their branching fractions.  Upper
vertex
$\bar D^0$ and $D^-$, and $b \rightarrow s/d\ g,\ \ s/d\ q \bar q$, are from
this analysis.  The branching fractions for the separate components making up
$b \rightarrow (c\ {\rm or}\ u)\ \bar c s^\prime$  are shown parenthetically. }
\vskip .5cm
\begin{tabular} {l  c  c c}
$b$ decay modes & \multicolumn{3}{c}{Branching fraction (\%)}        \\  \hline
$b \rightarrow (c\ {\rm or}\ u)\ e \nu$  &  $b_{SL}$ & & 10.5 $\pm$ 0.5  \\
$b \rightarrow (c\ {\rm or}\ u)\ \mu \nu$&  $b_{SL}$ & & 10.5 $\pm$ 0.5  \\
$b \rightarrow (c\ {\rm or}\ u)\ \tau \nu$& $0.25b_{SL}$& & 2.6 $\pm$ 0.1\\
$b \rightarrow (c\ {\rm or}\ u)\ \bar u d^\prime$&$r_{ud}b_{SL}$ & &
 42.0 $\pm$ 2.0 $\pm$ 4.2 \\  \hline
$b \rightarrow (c\ {\rm or}\ u)\ \bar c s^\prime$ &  & &  21.9 $\pm$ 3.7 \\
$\ \ \ \ \ \ \ \ \ D_s$          &                & (10.0 $\pm$ 2.7) &   \\
$\ \ \ \ \ \ \ \ \ (c \bar c)$   &                &   (3.0 $\pm$ 0.5) &   \\
\ \ \ \ \ \ \ \ \ baryons        &                &   (1.0 $\pm$ 0.6)  & \\
\ \ \ \ \ \ \ \ upper vertex $\bar D^0,D^-$ &   &   (7.9 $\pm$ 2.2) & \\  \hline
$b \rightarrow s/d\ g,\ s/d\ q \bar q$ &     &  & 0.2 $\pm$ 4.1  \\ \hline
\hline
TOTAL                           &           & & 87.7 $\pm$ 7.4    \\
\end{tabular}
\end{center}
\end{table}


\begin{references}
\bibitem{conjugate} Throughout, charge conjugation of all equations is implied,
i.e., $B \rightarrow \bar D X$ includes $\bar B \rightarrow D \bar X$.
\end{references}
\end{document}